\documentclass[sigconf, manuscript, nonacm]{acmart}

\renewcommand\footnotetextcopyrightpermission[1]{} 
\AtBeginDocument{%
  }

\usepackage{import}
\newcounter{rqcounter}
\newcommand\alias[2]{\expandafter\def\csname alias:#1\endcsname{#2}}
\newcommand\A[1]{\csname alias:#1\endcsname}

\usepackage{enumitem}

\definecolor{mygreen}{RGB}{189,178,70}
\definecolor{mypink}{RGB}{202,45,85}
\definecolor{myblue}{RGB}{0,57,92}

\usepackage{fdsymbol}
\usepackage{booktabs}
\usepackage{tabularx}
\usepackage{multirow}
\usepackage{comment}
\usepackage{graphicx}
\usepackage{url}
\usepackage{circledsteps}
\usepackage{tcolorbox}
\usepackage{caption}
\usepackage{subcaption}
\usepackage{colortbl}

\usepackage{longtable}
\usepackage{wasysym}

\usepackage[%
acronym
]{glossaries}

\newglossaryentry{log4j}{%
    name={Log4j},
    description={},
}
\newglossaryentry{solarwinds}{
    name={SolarWinds},
    description={},
}
\newglossaryentry{xz}{
    name={XZ Utils},
    description={},
}
\newglossaryentry{attack}{
    name={attack technique},
    description={},
    first={attack techniques}
}
\newglossaryentry{safeguard}{
    name={task},
    description={},
    first={tasks},
}
\newglossaryentry{starter}{
    name={\textit{starter kit}},
    description={},
}

\newacronym{cncf}{CNCF}{Cloud Native Computing Foundation}
\newacronym[first=Secure Software Development Framework (SSDF)]{ssdf}{SSDF}{Secure Software Development Framework}
\newacronym{psscrm}{P-SSCRM}{Proactive Software Supply Chain Risk Management}
\newacronym{slsa}{SLSA}{Supply-chain Levels for Software Artifacts}
\newacronym{eo}{EO}{Executive Order}
\newacronym{scorecard}{OpenSSF Scorecard}{OpenSSF Scorecard}
\newacronym{bsimm}{BSIMM}{Building Security In Maturity Model}
\newacronym{s2c2f}{S2C2F}{OpenSSF Secure Supply Chain Consumption Framework}
\newacronym{owaspSCVS}{OWASP SCVS}{OWASP Software Component Verification Standard}
\newacronym{cncfSSC}{CNCF SSC}{CNCF Software Supply Chain Best Practices}
\newacronym{attestation}{Self-Attestation}{Self-Attestation}
\newacronym{attck}{MITRE ATT\&CK}{MITRE Adversarial Tactics, Techniques and Common Knowledge}
\newacronym[first={Cyber Threat Intelligence (CTI)}]{cti}{CTI}{Cyber Threat Intelligence}
\newacronym{nist}{NIST}{National Institute of Standards and Technology}
\newacronym[firstplural=Common Vulnerabilities and Exposures (CVEs)]{cve}{CVE}{Common Vulnerabilities and Exposures}
\newacronym{cwe}{CWE}{Common Weakness Enumeration}
\newacronym{cisa}{CISA}{Cybersecurity and Infrastructure Security Agency}
\newacronym{nvd}{NVD}{National Vulnerability Database}
\newacronym{cna}{CNA}{CVE Numbering Authority}
\newacronym{llm}{LLM}{Large Language Models}
\newacronym[first=Software Bills of Materials (SBOMs)]{sbom}{SBOM}{Software Bill of Materials}
\newacronym{ttps}{TTPs}{Tactics, Techniques, and Procedures}
\newacronym{cot}{CoT}{Chain of Thought}
\newacronym{jndi}{JNDI}{Java Naming and Directory Interface}
\newacronym{poc}{PoC}{Proofs of Concept}
\newacronym{tp}{TP}{True Positive}
\newacronym{fp}{FP}{False Positive}
\newacronym{fn}{FN}{False Negative}
\newacronym[first=Attacks Under Study (AUS)]{aus}{AUS}{Attacks Under Study}

\newacronym{2fa}{2FA}{two-factor authentication}
\newacronym{bdfl}{BDFL}{Benevolent Dictator for Life}
\newacronym{capec}{CAPEC}{Common Attack Pattern Enumeration and Classification}
\newacronym{cra}{CRA}{Cyber Resilience Act}
\newacronym{ci}{CI}{continuous integration}
\newacronym{cicd}{CI/CD}{continuous integration and continuous delivery}
\newacronym{cipac}{CIPAC}{Critical Infrastructure Partnership Advisory Council}
\newacronym{cla}{CLA}{Contributor License Agreement}
\newacronym{cvss}{CVSS}{Common Vulnerability Scoring System}
\newacronym{dpo}{DPO}{Data Protection Officer}
\newacronym{enisa}{ENISA}{European Union Agency for Cybersecurity}
\newacronym{epss}{EPSS}{Exploit Prediction Scoring System}
\newacronym{esf}{ESF}{Enduring Security Framework}
\newacronym{gdpr}{GDPR}{General Data Protection Regulation}
\newacronym{irb}{IRB}{Institutional Review Board}
\newacronym{ngo}{NGO}{non-governmental organization}
\newacronym{ntia}{NTIA}{National Telecommunications and Information Administration}
\newacronym{openssf}{OpenSSF}{Open Source Security Foundation}
\newacronym{oss}{OSS}{open source software}
\newacronym{osc}{OSC}{open source component}
\newacronym{osp}{OSP}{open source project}
\newacronym{sat}{SAT}{static analysis tool}
\newacronym{satc}{SaTC}{Secure and Trustworthy Cyberspace}
\newacronym{sme}{SME}{small and medium enterprises}
\newacronym{ssc}{SSC}{software supply chain}
\newacronym{sso}{SSO}{single sign-on}
\newacronym{svm}{SVM}{Support Vector Machine}
\newacronym{tls}{TLS}{Transport Layer Security}
\newacronym{vcc}{VCC}{vulnerability-contributing commit}
\newacronym{vex}{VEX}{Vulnerability-Exploitability Exchange}

\usepackage{longtable}
\usepackage{booktabs}
\usepackage{caption}
\usepackage{graphicx}
\setcopyright{acmlicensed}
\copyrightyear{2025}
\acmYear{2025}
\acmDOI{XXXXXXX.XXXXXXX}
\acmConference[Conference acronym 'XX]{Make sure to enter the correct
  conference title from your rights confirmation email}{June 03--05,
  2018}{Woodstock, NY}

\acmISBN{978-1-4503-XXXX-X/18/06}

\begin{document}

\title{Your ATs to Ts: MITRE ATT\&CK Attack Technique to P-SSCRM Task Mapping}
\subtitle{Version 1.1}

\author{Sivana Hamer}
\email{sahamer@ncsu.edu}
\affiliation{%
  \institution{North Carolina State University}
  \city{Raleigh}
  \state{NC}
  \country{US}
}

\author{Jacob Bowen}
\email{jebowen2@ncsu.edu}
\affiliation{%
  \institution{North Carolina State University}
  \city{Raleigh}
  \state{NC}
  \country{US}
}

\author{Md Nazmul Haque}
\email{mhaque4@ncsu.edu}
\affiliation{%
  \institution{North Carolina State University}
  \city{Raleigh}
  \state{NC}
  \country{US}
}

\author{Brandon Wroblewski}
\email{bnwroble@ncsu.edu}
\affiliation{%
  \institution{North Carolina State University}
  \city{Raleigh}
  \state{NC}
  \country{US}
}

\author{Chris Madden}
\email{chris.madden@yahooinc.com}
\affiliation{%
  \institution{Yahoo}
  \city{Dublin}
  \country{Ireland}
}

\author{Laurie Williams}
\email{lawilli3@ncsu.edu}
\affiliation{%
  \institution{North Carolina State University}
  \city{Raleigh}
  \state{NC}
  \country{US}
}

\maketitle

\section*{Revision History}
~\autoref{tab:revision_history} summarizes the changes incorporated into each release of the MITRE ATT\&CK attack techniques to P-SSCRM tasks mapping. Version~1.0 is the first released version. Version~1.1 is a minor update that extends the mapping with no existing V~1.1 mappings removed or modified.

\begin{table}[htbp]
\centering
\small
\caption{Revision history of the MITRE ATT\&CK attack techniques to P-SSCRM tasks mapping.}
\label{tab:revision_history}
\resizebox{\columnwidth}{!}{%
\begin{tabular}{@{}p{1.8cm}p{1.6cm}p{11cm}@{}}
\toprule
\textbf{Date} & \textbf{Version} & \textbf{Description of Changes} \\
\midrule
\textit{24.07.2025} & 1.0 & Initial release of the MITRE ATT\&CK attack techniques (v16.1) to P-SSCRM tasks (v1.01) mapping. Mappings were selected by requiring agreement among at least three of the four triangulation strategies described in Sec.~\ref{SEC:Procedure}. The mapping covers 94 out of 203 MITRE ATT\&CK attack techniques and 42 out of 73 P-SSCRM tasks.\\
\addlinespace[3pt]
\textit{03.08.2026} & 1.1 & Minor update of the MITRE ATT\&CK attack techniques (v16.1) to P-SSCRM tasks (v1.01) mapping.
The mapping was manually extended to cover 42 MITRE ATT\&CK techniques observed in the 106 analyzed CTI reports but with no mappings found in Version~1.0 described in Sec.~\ref{SEC:Procedure}.
The mapping covers 136 of 203 unique MITRE ATT\&CK attack techniques and 43 out of 73 P-SSCRM tasks.\\
\bottomrule
\end{tabular}%
}
\end{table}

\noindent\textbf{Data availability}.
The current version of the mapping is provided in \autoref{tab:mitre-psscrm-mapping}.
Machine-readable versions of the mappings, with additional metadata on how the data was collected, are available at \url{https://github.com/p-sscrm/ats-to-ts}.
 
\section{Introduction and Background}
High-profile incidents -- affecting SolarWinds, Log4j, and XZ Utils-- have increased industry and research attention to software supply chain security. 
Software supply chain attacks occur when vulnerabilities are maliciously or accidentally introduced into dependencies to compromise dependents.
Software supply chain frameworks have compiled tasks of what software organizations should or have to do to reduce risk.
For example, the US \gls{nist} \glsfirst{ssdf}~\cite{souppaya2022secure} describes different tasks such as providing role-based training, securely archiving the data for each software release, and analyzing vulnerabilities to determine root causes.
The \gls{psscrm} framework unifies 10 notable software supply chain frameworks into 73 tasks~\cite{williams2024proactive}.

However, no mapping between \gls{attack} and \gls{safeguard} exists for the software supply chain.
Therefore, we do not know how different tasks mitigate the attack techniques of software supply chain attacks.
In this document, we present our mapping between \gls{attck}~\cite{mitreATTACK} \glspl{attack} to \gls{psscrm} \glspl{safeguard}~\cite{williams2024proactive} which originated from our research paper~\cite{hamer2025closingchainreducerisk}. 
\textbf{Because each P-SSCRM task is mapped to one or more tasks from the 10 frameworks, the mapping we provide is also a mapping between \gls{attck} and other prominent government and industry frameworks.}
For example, through our mapping, the US \gls{nist} \glsfirst{ssdf} tasks PO.5.1 and PO.5.2 are mapped in \gls{psscrm} to environmental separation (E.3.2) and, based on our mapping, mitigates the attack technique exploiting public-facing applications (T1190).
A machine-readable format is also available on GitHub~\cite{githubRepo}.
In this manuscript, we describe related frameworks (Sec.~\ref{SEC:Frameworks}, explain relevant terminology (Sec.~\ref{SEC:Terminology}) explain how we gathered the mapping (Sec.~\ref{SEC:Procedure}), how to use the mapping (Sec.~\ref{SEC:How}), and who can use the mapping (Sec.~\ref{SEC:Who}).

\subsection{Related Frameworks}~\label{SEC:Frameworks}

Below are the related frameworks used throughout this document.

\begin{itemize}
    \item \textbf{MITRE ATT\&CK:} The framework provides a taxonomy of \gls{ttps} used by adversaries across the attack life cycle~\cite{mitreATTACK}.
    We use the enterprise version \gls{attck} (v16.1), which has 14 tactics, 203 techniques, and 453 sub-techniques.
    \item \textbf{P-SSCRM:} The framework maps bi-directionally equivalent tasks described in ten prominent referenced risk management software supply chain frameworks~\cite{williams2024proactive}. The frameworks within P-SSCRM are \gls{eo} 14028~\cite{EO_2021}, \gls{ssdf}~\cite{souppaya2022secure}, NIST 800 161~\cite{NIST_800161}, Self attestation~\cite{Attestation}, \gls{bsimm}~\cite{BSIMM}, \gls{slsa}~\cite{SLSA}, \gls{s2c2f}~\cite{Microsoft_framework}, \gls{owaspSCVS}~\cite{scvs}, and \gls{scorecard}~\cite{Scorecard}. The mappings between P-SSCRM and the frameworks, with the respective framework-to-framework mappings, can be found in the P-SSCRM documentation~\cite{williams2024proactive}.
    We use \gls{psscrm} version (v1.01) that contains 4 groups, 15 practices, and 73 tasks.
\end{itemize}

\subsection{Terminology \& Definitions}\label{SEC:Terminology}

Below are relevant terminology and definitions used throughout this document:

\begin{itemize}
    \item \textbf{Mapping:} Indicates that a concept, unit of knowledge uniquely characterized by a combination of characteristics, relates to another concept~\cite{mappingNIST}.
    \item \textbf{Attack technique:} 
    The actions used by an adversary to achieve a goal. We describe attack techniques using \gls{attck}~\cite{mitreATTACK}.
    Each attack technique has an identifier of the form TXXXX.
    An example of an attack technique is valid accounts (T1078).
    Each attack technique is mapped to at least one tactic, which represents the goal of the attack technique.
    Additionally, certain attack techniques have sub-techniques, specific descriptions of an attack technique. 
    \item \textbf{Task:} 
    A lower-level action or effort to secure the software application and reduce security risk. 
    We characterize software supply chain tasks through P-SSCRM~\cite{williams2024proactive}.
    Each task has an identifier of the form X.Y.Z. 
    An example of a task is updating your dependencies (P.5.2). 
    Each task has an associated practice (mid-level object) and group (high-level objective). 
    For example, managing vulnerable components and containers (P.5) is a practice, and the product (P) is the group.
\end{itemize}

\subsection{The MITRE ATT\&CK to P-SSCRM Mapping Process}~\label{SEC:Procedure}

In the following, we describe how each version of the mapping was constructed. 
In \textit{version 1.0} (Sec.~\ref {SEC:V1.0}), we created the initial mapping, which we later extended in \textit{Version 1.1} (Sec.~\ref{SEC:V1.1}).

\subsubsection{Version 1.0}\label{SEC:V1.0}
Mapping cyber-security frameworks is subjective~\cite{mappingNIST}, thus gathering where opinions converge is needed.
Triangulation is a research technique that utilizes multiple approaches to enhance the interpretation of findings~\cite{thurmond2001point}.
In our case, we use methodological approaches using four independent, systematic, and scalable strategies to triangulate results between strategies. 
We then select the \gls{attck} \gls{attack} to \gls{psscrm} \gls{safeguard} mappings from the strategies. In the following paragraphs, we detail our strategies and triangulation approach from our research paper~\cite{hamer2025closingchainreducerisk}.

\begin{itemize}
    \item \textbf{Transitive mapping (M1):}
    Although there is no direct \gls{attck} attack technique to \gls{psscrm} task mapping, transitive mappings between frameworks can be found.
    For example, we can map \gls{attck} attack techniques $\leftrightarrow$ \gls{nist} SP 800 53  $\leftrightarrow$ \gls{nist} \gls{ssdf} $\leftrightarrow$ \gls{psscrm} tasks.
     For example, from the transitive mapping, we get the mapping between unsecured credentials (T1552) and automated security scanning tools (P.4.2) (T1552 $\leftrightarrow$ SA-15 $\leftrightarrow$ PO.3.2 $\leftrightarrow$ P.4.2).
    Hence, we constructed a dataset to leverage the transitive mappings.
    We start by finding the mappings to \gls{attck} and the ten referenced frameworks.
    We iteratively searched through web pages, Google searches, and documentation for mappings from and to the frameworks and found 183 framework-to-framework mappings. 

    We used the \textit{networkx} package in \textit{Python} to find all simple paths~\cite{networkx} between \gls{attck} and \gls{psscrm} to find paths with no repeated frameworks.
    We set a cutoff value for the path length as the number of simple paths increases exponentially.
    We discarded superset simple paths as we wanted the shortest simple paths. 
    For example, we discard the path \gls{attck} $\leftrightarrow$ \gls{nist} SP 800 53  $\leftrightarrow$ \gls{nist} \gls{ssdf}  $\leftrightarrow$ \gls{nist} 800 161 $\leftrightarrow$ \gls{psscrm} as the simple path was a superset of our prior example.
    As more than $99\%$ of simple paths with a length above $7$ superset another shorter path, we selected $10$ as a conservative threshold.
    We remained with 33 frameworks and 78 framework-to-framework mappings that generated 125 simple paths.
    We then collected the data, cleaned up inconsistencies, and found 17 simple paths with data.
    For the 108 simple paths without data, we manually checked the paths and found no data to create a transitive path in most cases (101 paths).
    For example, although a simple path exists between \gls{attck} $\leftrightarrow$ OWASP Community Attacks $\leftrightarrow$ OWASP Cheat Sheets, no attack technique maps to an OWASP cheat sheet.
    Other reasons were differences in the data granularity (5 paths) and no end-to-end transitive data (2 paths).

    \item \textbf{LLM mapping (M2):} 
    We leveraged \gls{llm} capabilities to find patterns in data to create attack technique to task mappings.
    We tested three models: ChatGPT gpt-4o-mini~\cite{chatgpt}, VertexAI with gemini-1.5-pro~\cite{vertexAI}, and Claude 3.5 Sonnet 2024-10-22~\cite{claude}.
    We selected the models as they provided APIs used in prior work~\cite{zahan2024shifting, bae2024enhancing} or included grounded models~\cite{grounded}.
    To reduce bias in our prompt, available in supplemental material~\cite{supplementalLink}, we base the wording using the US NIST's cybersecurity mapping standard~\cite{mappingNIST} and the frameworks we are mapping~\cite{mitreATTACK,williams2024proactive} using a \gls{cot} prompt inspired from prior work~\cite{dunlap2024pairing}.
    We ask to find bi-directionally supportive relationships where a task mitigates an attack technique.
    We also ask the prompt to produce a binary answer~\cite{tai2024examination} for a pair of attack technique to task mapping at a time to increase accuracy.
    For example, we ask the \gls{llm} if the task of producing attestation (G.1.3) mitigates the \gls{attack} masquerading (T1036), from which we get there is a mapping.
    
    We choose the \gls{llm} based on a sample of attack technique to task mappings.  
    We stratified our sample~\cite{baltes2022sampling} based on the combinations of \gls{attck} tactics and \gls{psscrm} groups to ensure we get pairs of each group.
    We then randomly selected, proportional to the size of the number of pairings, 150 samples to have at least one pair for each strata.
    For example, without the new attack techniques, there are 44 attack techniques in the defense evasion (TA0005) tactic and 23 tasks in the governance (G) group leading to $44 \times 23 = 1,012$ possible pairings.
    Considering some attack techniques belong to multiple tactics, the number of possible pairings across all groups is $73 \times  236 =17,228$.
    The number of pairs for the strata by proportional sampling is $\lceil\frac{1,012}{17,228} \times 150 \rceil = 9$.
    We tested both zero-shot and one-shot versions of the prompt in each model, generating 900 \gls{llm} responses.
    Two authors independently mapped if the task mitigated the attack technique for the 150 samples and negotiated disagreements. 
    We selected ChatGPT GPT-4o-mini as the model with a zero-shot prompt as: 
    (a) zero-shot outperformed one-shot, 
    (b) 82\% of the pairs were the same as the disagreement-resolved sample, and
    (c) the cost was only 3\% of the other models.
    
    \item \textbf{Framework mapping (M3):} 
    The \gls{attck} framework also contains general software security mitigations, which we call tasks in this paper, that map to attack techniques.
    Thus, we map \gls{attck} and \gls{psscrm} tasks and leverage the existing mapping (\gls{attck} attack techniques $\leftrightarrow$ \gls{attck} tasks $\leftrightarrow$ \gls{psscrm} tasks).
    For example, we mapped the \gls{attck} mitigation code signing (M1045) with the \gls{psscrm} task requiring signed commits (P.3.3). 
    We followed the NIST mapping standard~\cite{mappingNIST}, mapping set relationships between \gls{attck} and \gls{psscrm} tasks, as there were few bi-directionally equivalent tasks.
    The first author created the  \gls{attck} and \gls{psscrm} task set mapping, which was then reviewed by the second author.

    \item \textbf{Report mapping (M4):} 
    \label{sec:report_mapping}
    We qualitatively analyzed 106 \gls{cti} reports of three software supply chain attacks (SolarWinds, Log4j, and XZ Utils), from which we manually mapped mitigated \gls{attack} to \gls{safeguard}. 
    For example, the quote \textit{``as part of our [\gls{solarwinds}] response to the SUNBURST vulnerability, the code-signing certificate used by SolarWinds to sign the affected software versions was revoked...''} is mapped to the subvert trust controls (T1533) and trusted relationship (T1199) \glspl{attack} with the software release integrity (P.1.2) \gls{safeguard}.
    When available, we also extracted recommended tasks, which we defined as suggestions for mitigating the attacks.
    For example, a task recommendation in \gls{solarwinds} is \textit{``organizations need to harden their build environments against attackers''} is mapped to the subvert trust controls (T1533) attack technique with CI/CD hosting and automation (E.2.4) task.
    We found 14 unique non-adversarial events in \gls{solarwinds}, 58 in \gls{log4j}, and 53 in \gls{xz}.
    Meanwhile, we found 653 task recommendations in \gls{solarwinds}, 791 in \gls{log4j}, and 180 in \gls{xz}.
\end{itemize}

We found 5,766 candidate attack techniques to tasks mappings from the four strategies, which resulted in 4,458 unique candidate mappings after removing duplicates.
We calculated Krippendorff's alpha to determine the agreement between strategies, resulting in a value of 0.1, indicating slight agreement~\cite{gonzalez2023reliability}.
Hence, strategies are not substituting for each other, aligning with methodological triangulation~\cite{thurmond2001point}. 
We thus filter the attack technique to task mappings to select only agreed-upon mappings.
We selected three strategies as our threshold to find where opinions converged while allowing us to find some differences between strategies. 
Therefore, we selected 251 mappings between MITRE ATT\&CK attack techniques to P-SSCRM tasks.
The mapping covered 94 out of 203 MITRE ATT\&CK attack techniques and 42 out of 73 P-SSCRM tasks.

\subsubsection{Version 1.1}\label{SEC:V1.1}

We manually extended the mapping to cover MITRE ATT\&CK attack techniques for which no agreed-upon mapping exists (109 out of 203 MITRE ATT\&CK attack techniques).
We focused on 42 MITRE ATT\&CK techniques leveraged by adversaries in the SolarWinds, Log4j, and XZ Utils campaigns from our initial paper~\cite{hamer2025closingchainreducerisk}.
We used our 4,458 unique candidate mappings from Sec.~\ref{SEC:V1.0} to increase the likelihood of finding actual mappings, but excluded the 251 agreed-upon mappings from v1.0.
As the manual effort of checking the remaining 4,207 candidate mappings was considerable, we only mapped candidate mappings: (i) found by the report mapping strategy (M4) as the mappings were manually curated, and (ii) found by at least one other strategy (M1, M2, or M3).
We remained with 437 MITRE ATT\&CK attack techniques to P-SSCRM candidate tasks mappings.
Two researchers then independently evaluate the remaining 437 candidate mappings.
Each research study referenced the MITRE ATT\&CK attack technique descriptions and tasks, and the P-SSCRM task objectives, descriptions, and assessment questions.
We calculated Cohen's Kappa to assess agreement between the two researchers, resulting in $\kappa = 0.85$ with an almost perfect agreement~\cite{gonzalez2023reliability}.
The remaining 26 disagreements were resolved by another author and discussed with one of the two authors involved in the 437 candidate task mappings in v1.1.
We therefore curated an additional 79 mappings between MITRE ATT\&CK attack techniques to P-SSCRM tasks.
The mapping covered 94 out of 203 MITRE ATT\&CK attack techniques and 42 out of 73 P-SSCRM tasks.

\subsection{How Should I Use the MITRE ATT\&CK to P-SSCRM Mapping?}\label{SEC:How}

The two main use cases for using the MITRE ATT\&CK to P-SSCRM mapping: 

\begin{itemize}
    \item By \textbf{leveraging} the \gls{attck} to \gls{psscrm} mapping to collect, given an \gls{attack}, what are agreed upon software supply chain \glspl{safeguard} that reduce security risk. 
    \Glspl{attack} can be gathered manually or using automated approaches.
    Additionally, the \glspl{attack} can be from both private and public reports.
    \item By \textbf{producing} new mappings between frameworks using our procedure to collect the convergence of different strategies (Section~\ref{SEC:Procedure}). 
    The procedure can be used with the existing strategies or extended with new strategies.
\end{itemize}

\subsection{Who Can Use the MITRE ATT\&CK to P-SSCRM Mapping?}\label{SEC:Who}

The mapping is appropriate to be used for anyone interested in understanding the relationship between attack techniques and tasks in the software supply chain.
The following are five user stories of how the mapping can be used:

\begin{itemize}
    \item As a \textbf{software organization}, I want to know which \glspl{safeguard} mitigate which \glspl{attack}, so that I can prioritize tasks. 
    \item As a \textbf{cyber-threat analyst}, I want to know which \glspl{safeguard} mitigate which \glspl{attack}, so that I can recommend \glspl{safeguard} given an attack.
    \item As an \textbf{auditor}, I want to know which \glspl{safeguard} mitigate which \glspl{attack}, to know which \glspl{safeguard} should I consider during an audit.
    \item As a \textbf{framework author}, I want to know which \glspl{safeguard} mitigate which \glspl{attack}, to validate the \gls{safeguard} in my framework given the goals of the framework.
    \item As an \textbf{framework mapping author}, I want to have a procedure to systematically map frameworks, to produce new mappings.
\end{itemize}

\section{MITRE ATT\&CK attack techniques to P-SSCRM tasks mappings}\label{SEC:Mapping}
\autoref{tab:mitre-psscrm-mapping} lists \gls{attck} \gls{attack} and their corresponding P-SSCRM tasks. We found 136 out of 203  \gls{attck} \gls{attack} techniques and 43 out of 73 \gls{psscrm} tasks.
\begin{longtable}{l p{9cm}}
\caption{MITRE ATT\&CK Techniques Mapped to PSSCRM Tasks}
\label{tab:mitre-psscrm-mapping} \\
\toprule
\textbf{MITRE Technique} & \textbf{PSSCRM Task} \\
\midrule
\endfirsthead
\multicolumn{2}{c}%
{{\tablename\ \thetable{} -- continued from previous page}} \\
\toprule
\textbf{MITRE Technique} & \textbf{PSSCRM Task} \\
\midrule
\endhead
\bottomrule
\multicolumn{2}{r}{Continued on next page} \\
\endfoot
\bottomrule
\endlastfoot
T1001 & E.3.7 \\
T1003 & D.2.1, E.3.3 \\
T1005 & G.2.6 \\
T1008 & E.3.7 \\
T1021 & D.2.1, E.3.3, E.3.7 \\
T1025 & G.2.6 \\
T1027 & D.2.1, E.3.11 \\
T1036 & D.2.1, E.3.3, E.3.11 \\
T1040 & E.3.2, E.3.4 \\
T1041 & D.2.1, E.3.7, G.2.6 \\
T1046 & E.3.7, P.2.1 \\
T1047 & E.3.3, E.3.7 \\
T1048 & E.3.7, G.2.6 \\
T1052 & G.2.6 \\
T1059 & D.2.1, E.3.3, E.3.6 \\
T1068 & D.2.1, E.2.5, E.3.3, G.4.3, P.5.2 \\
T1070 & D.2.1 \\
T1071 & D.2.1, E.3.7 \\
T1072 & E.1.5, E.2.5, E.3.2, E.3.3, E.3.7, G.5.2, P.5.2 \\
T1078 & D.2.1, E.1.3, E.3.2, E.3.3, E.3.8, E.3.9, G.4.1 \\
T1080 & D.2.1, E.3.3, E.3.7, E.3.11 \\
T1090 & E.3.7 \\
T1095 & E.3.7 \\
T1098 & E.1.3, E.3.3, E.3.6, E.3.11 \\
T1102 & E.3.7 \\
T1104 & E.3.7 \\
T1105 & D.2.1, E.3.7 \\
T1110 & E.3.3 \\
T1114 & E.3.4, E.3.6 \\
T1127 & P.2.1 \\
T1133 & E.3.7 \\
T1134 & E.3.3 \\
T1136 & E.3.3 \\
T1176 & D.2.1, E.3.6, E.3.7 \\
T1185 & E.3.3 \\
T1187 & E.3.7 \\
T1189 & D.2.1, E.3.7 \\
T1190 & D.1.2, D.1.3, D.2.1, E.2.4, E.2.5, E.3.2, E.3.3, E.3.7, G.2.5, G.4.1, G.5.1, P.4.3, P.4.4, P.5.2, D.1.7, E.3.11 \\
T1195 & D.1.2, D.1.3, D.1.4, D.1.6, D.2.1, G.1.4, G.1.5, G.3.1, G.4.1, G.4.3, G.5.1, P.2.1, P.3.4, P.4.1, P.4.3, P.4.5, P.5.1, P.5.2, D.1.7, E.3.11, G.5.5 \\
T1199 & E.3.3, E.3.7, G.5.1, P.2.1, D.1.7, E.3.11, G.5.5 \\
T1203 & D.2.1, E.3.7 \\
T1204 & E.3.7 \\
T1205 & E.3.7 \\
T1210 & D.2.1, E.3.3, E.3.7, G.4.3, P.2.1, P.4.3, P.5.2 \\
T1211 & D.2.1, E.2.5, G.4.3, P.5.2 \\
T1212 & D.2.1, E.2.5, E.3.7, G.4.3, P.5.2 \\
T1213 & D.2.1, E.3.3, E.3.4, G.2.6 \\
T1218 & D.2.1, E.3.3 \\
T1219 & E.3.7 \\
T1221 & E.3.7, P.2.1, E.3.11 \\
T1222 & E.3.3 \\
T1484 & D.2.1, E.3.3, E.3.6 \\
T1489 & E.3.3 \\
T1505 & E.3.3, E.3.6, P.2.1, P.4.5 \\
T1525 & E.3.3, E.3.6 \\
T1528 & D.2.1, E.3.3 \\
T1530 & E.3.3, E.3.4, G.2.6 \\
T1537 & E.3.3, E.3.7 \\
T1538 & E.3.3 \\
T1543 & D.2.1, E.3.6 \\
T1548 & D.2.1, E.3.3, E.3.6 \\
T1550 & D.2.1, E.3.3, E.3.6, G.3.4 \\
T1552 & D.2.1, E.3.3, E.3.4, E.3.6, E.3.7, E.3.8, G.2.6 \\
T1553 & D.1.2, D.1.4, E.3.3, P.2.1, P.4.5, P.5.2 \\
T1554 & P.3.3 \\
T1555 & D.2.1, E.3.3 \\
T1556 & D.2.1, E.3.3, E.3.6 \\
T1557 & E.3.2, E.3.4, E.3.7 \\
T1558 & D.2.1, E.3.3, E.3.8 \\
T1559 & P.2.1 \\
T1562 & D.1.2, D.2.1, E.3.6 \\
T1563 & E.3.2, E.3.3 \\
T1565 & E.3.3, E.3.4, G.2.6 \\
T1566 & E.3.7 \\
T1567 & E.3.7 \\
T1568 & D.2.1, E.3.7 \\
T1569 & E.3.3 \\
T1570 & E.3.7 \\
T1571 & E.3.7 \\
T1572 & E.3.7 \\
T1573 & D.2.1, E.3.7 \\
T1574 & D.2.1, E.3.6 \\
T1578 & E.3.3 \\
T1599 & E.3.7 \\
T1602 & E.3.4, E.3.7 \\
T1606 & D.2.1, E.3.3, G.3.4 \\
T1609 & E.3.3, P.2.1 \\
T1610 & E.3.3, E.3.6, E.3.7 \\
T1611 & E.3.7, P.2.1 \\
T1612 & E.3.6, P.4.5 \\
T1621 & E.3.3 \\
T1649 & E.3.3 \\
T1651 & E.3.3 \\
T1659 & E.3.4, E.3.7 \\
T1564 & E.3.11 \\
T1584 & E.3.11 \\
T1608 & E.3.11 \\
T1007 & P.2.1 \\
T1012 & D.2.1 \\
T1014 & E.2.6, E.3.10, E.3.6 \\
T1016 & D.2.1 \\
T1018 & D.2.1 \\
T1033 & P.2.1 \\
T1049 & D.2.1 \\
T1053 & D.2.1, D.2.2, E.3.3, E.3.6, G.3.4, P.2.1 \\
T1055 & D.2.1, D.2.2, E.3.10, E.3.3, E.3.7, P.4.4 \\
T1056 & E.1.3, E.3.1, G.3.4 \\
T1057 & D.2.1 \\
T1082 & D.2.1 \\
T1083 & D.2.1 \\
T1112 & E.3.3, E.3.6 \\
T1119 & D.2.1, E.3.4, G.2.6 \\
T1129 & D.2.1 \\
T1132 & D.2.1, E.3.7 \\
T1140 & D.2.1 \\
T1482 & D.2.1, E.3.2 \\
T1486 & D.2.1, E.3.3 \\
T1490 & D.2.1, E.3.3, E.3.6 \\
T1496 & D.2.1, E.3.3 \\
T1497 & D.2.1 \\
T1499 & D.1.2, D.2.1, E.3.7 \\
T1518 & D.2.1 \\
T1546 & D.2.1, D.2.2, E.3.1, E.3.3 \\
T1547 & D.2.1 \\
T1560 & D.2.1, G.3.4 \\
T1585 & E.3.3 \\
T1586 & D.2.1, E.1.3, E.1.4, E.3.1, E.3.3, E.3.9, G.3.4 \\
T1587 & G.4.3, G.5.1, P.2.1 \\
T1588 & G.4.3 \\
T1589 & G.3.4 \\
T1591 & G.3.4 \\
T1593 & G.4.1 \\
T1595 & D.2.1, G.4.3 \\
T1622 & D.2.1 \\
T1656 & G.4.1, G.4.3 \\
T1665 & D.2.1, E.3.7 \\
\end{longtable}

\begin{acks}
The work was supported and funded by the National Science Foundation Grant No. 2207008 and the North Carolina State University Goodnight Doctoral Fellowship. 
Any opinions expressed in this material are those of the authors and do not necessarily reflect the views of any of the funding organizations. 
We profoundly thank the Realsearch, WSPR, and S3C2 research group members, visitors, and collaborators for their support and feedback.
\end{acks}

\bibliographystyle{ACM-Reference-Format}
\bibliography{references}

\end{document}